\documentclass{aa} 

\newcommand\nice[1]{#1}    \newcommand\subm[1]{}   

\nice{\newcommand\dosingle[1]{#1}  \newcommand\dodouble[1]{ }} 
\subm{\newcommand\dosingle[1]{ }   \newcommand\dodouble[1]{#1}} 

\newcommand\bfrevisedversion[1]{ {#1} }

\usepackage{graphicx}  

\usepackage{amsfonts}
\usepackage{mathpi}
\usepackage{url}

\def\SSS{Sect.~}

\nice{ \newcommand\mycaptionfont{\protect\footnotesize} }

\nice{ \textheight 240mm \headheight 0pt \topmargin -2.1cm }

\def\gtapprox{\,\lower.6ex\hbox{$\buildrel >\over \sim$} \, }
\def\ltapprox{\,\lower.6ex\hbox{$\buildrel <\over \sim$} \, }
\def\propapprox{\,\lower.6ex\hbox{$\buildrel \propto\over \sim$} \, }

\def\arcs{\ifmmode {'' }\else $'' $\fi}     
\def\arcm{\ifmmode {' }\else $' $\fi}       

\def\fr7{7$ \hskip -0.9ex \vrule height0.8ex width0.8ex depth-0.73ex
                                                     \hskip0.1ex$}
\def\frtoday{Le\space\number\day\space\ifcase\month\or
  janvier\or f\'evrier\or mars\or avril\or mai\or juin\or
  juillet\or ao\^ut\or septembre\or octobre\or novembre\or 
d\'ecembre\fi\space \number\year}

\newcommand\hMpc{\mbox{$h^{-1}$ Mpc}}
\newcommand\hGpc{\mbox{$h^{-1}$ Gpc}}

\newcommand\Omm{\Omega_{\mbox{\rm \small m}}}
\newcommand\Omtot{\Omega_{\mbox{\rm \small tot}}}

\newcommand\bmath[1]{{\bf #1}}  

\title{A constraint
on any topological lensing hypothesis in the spherical case: it
must be a root of the identity}

\author{Boudewijn F. Roukema} 

\institute{Toru\'n Centre for Astronomy, N. Copernicus University,
ul. Gagarina 11, PL-87-100 Toru\'n, Poland
}

\date{\frtoday}

\titlerunning{Topological lensing in ${{}{S}}^3$ as $\sqrt[n]{I}$}
\authorrunning{Roukema}

\abstract{Three-dimensional catalogues of objects at cosmological
distances can potentially yield candidate 
topologically lensed pairs of sets of objects, which would
be a sign of the global topology of the Universe. In the spherical
case (i.e. if curvature is positive), a necessary condition, which
does not exist for either null or negative curvature, can
be used to falsify such hypotheses, {\em without needing to loop through
a list of individual spherical 3-manifolds}. This condition is that the
isometry between the two sets of objects
must be a root of the identity isometry in the covering space 
${{}{S}}^3$.
This enables numerical falsification of topological lensing hypotheses 
without needing to assume any particular spherical 3-manifold.
By embedding ${S}^3$ in euclidean 4-space, $\mathbb{R}^4$,
this condition can be expressed as the requirement that 
$M^n = I$ for an integer $n$, where $M$ is the matrix representation of the 
hypothesised \bfrevisedversion{topological} 
lensing isometry and $I$ is the identity.
Moreover, this test becomes even simpler with the requirement that 
the two rotation angles,  $\theta,\phi$,  corresponding to the given isometry,
satisfy $   {2\pi \over \theta},   {2\pi \over \phi} \in {\mathbb{Z}}$.
The calculation of this test involves finding the two eigenplanes of
the matrix $M$.
A GNU General Public Licence numerical package, called {\tt eigenplane},
is made available for finding the rotation angles 
and eigenplanes of an arbitrary isometry $M$ of $S^3$.
}

\begin{document}

\maketitle

\dodouble{\clearpage} 


\def\fPSHdeltaz{
\begin{figure}[ht]
\centering
\includegraphics[scale=0.53]{gmod_hist.ps}
\caption[]{ \mycaptionfont
PSH (pair separation histogram) 
for the objects in our sample. The only filter applied is 
that the redshifts to
both objects in a pair must be very similar --- 
\SSS\protect\ref{s-dzfilter}. The tolerance
on errors in the determination of redshifts is 0.5\%
difference in redshifts in a pair. The horizontal axis shows separations of
objects in pairs in units 
of {\hMpc}. The vertical axis shows the number of pairs in
a separation bin equal to 1{\hMpc}.}
\label{f-PSHdeltaz} 
\end{figure}
} 

\newcommand\fpairtypes{
\begin{figure}[ht]
\centering
\includegraphics[width=8cm]{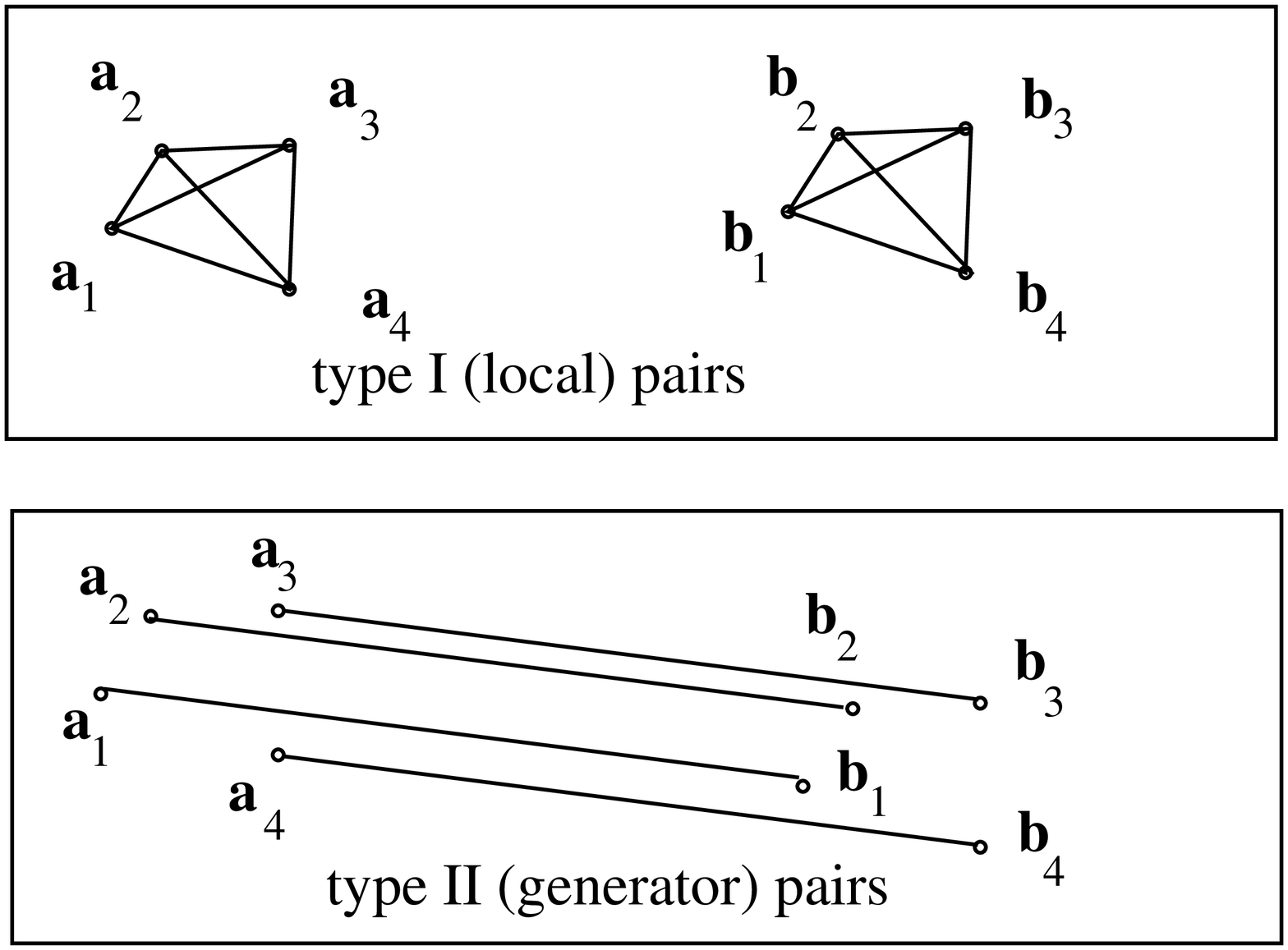}
\caption[]{
\mycaptionfont 
{\em Upper plot:} 
\bfrevisedversion{Example 
of $N=2$ images of a fundamental domain, 
where object $a_i$ is mapped to object $b_i$, i.e. $g(a_i)=b_i$.}
Each ``local'' (comoving spatial) geodesic between ``local'' objects should
occur \bfrevisedversion{$N=2$ times, i.e. twice in this example}. 
This frequency of occurrence is higher than for a Poisson 
distribution, in which any geodesic joining two objects should occur only once,
not \bfrevisedversion{$N$ times}.
{\em Lower plot:}
\bfrevisedversion{
Between the same two images of a fundamental domain, the geodesic connecting 
corresponding objects is a generator $g$ of the (physical) 
3-manifold from the (apparent) covering space.} 
\bfrevisedversion{If $g$ is a Clifford translation, then
it occurs many times with (ideally) exactly the same length,}
again more frequently than that of a Poisson 
distribution. 
}
\label{f-pairtypes}
\end{figure}
} 


\section{Introduction}
Although interest in cosmic topology is just over a century old
\nocite{Schw00,Schw98}({Schwarzschild} 1900, 1998), much interest has recently developed 
in trying to find clues to 
the topology of the Universe, in particular due to WMAP
cosmic microwave background observations
\nocite{WMAPSpergel}({Spergel} {et~al.} 2003) 
which have been found by many to have unusual statistical 
properties on the largest angular scales 
\nocite{WMAPSpergel,WMAPTegmarkFor,WMAPChiang,WMAPmultipol}(e.g.  {Spergel} {et~al.} 2003; {Tegmark} {et~al.} 2003; {Chiang} {et~al.} 2003; {Copi} {et~al.} 2004),
even though one study
across a wide range of parameter space 
failed to find any significant signal \nocite{CSSK03}({Cornish} {et~al.} 2004). 

Moreover, both the WMAP observations and 
supernovae Ia redshift-magnitude tests 
\nocite{SCP9812}(e.g.  {Perlmutter} {et~al.} 1999) suggest that the Universe is spherical, 
but with a radius of curvature at least as large as the horizon radius.
Indeed, the analysis of \nocite{ChouPaddy03}{Choudhury} \& {Padmanabhan} (2003) rejects the flat universe hypothesis
in favour of a spherical model at either $97\%$ or $90\%$ confidence
levels, depending on which data set of type Ia supernovae is analysed.

In addition, some analyses \nocite{LumNat03,RLCMB04}({Luminet} {et~al.} 2003; {Roukema} {et~al.} 2004) suggest that the
spatial comoving section of the Universe is a Poincar\'e dodecahedral 
\bfrevisedversion{space,} 
which is a 3-manifold with ${{}{S}}^3$ as its covering space, 
i.e. curvature is positive.

This type of hypothesis implies tight constraints on the curvature 
parameters: the non-relativistic matter density parameter, $\Omm,$ 
the cosmological constant (or quintessence constant) 
$\Omega_\Lambda$, and their sum, $\Omtot = \Omm + \Omega_\Lambda.$

While much interest is presently focussing on microwave background
analyses,  
only a relatively small error in the present estimates of
$\Omtot$ is required in order for topological lensing of sub-microwave
background objects, such as quasars, to be detectable.

For example, in the right-hand plot of fig.~15 of \nocite{GausSph01}{Gausmann} {et~al.} (2001),
it was shown that for 
\bfrevisedversion{the Poincar\'e dodecahedral space and}
$(\Omm=0.35,\Omega_\Lambda=0.75)$, i.e. 
for $\Omtot = 1.1$, a spike in
the pair separation histogram (PSH) would be expected for a catalogue
containing objects in the $1 < z < 3$ redshift range --- the range of
most interest for large quasar catalogues.

Is $\Omtot=1.1$ reasonable? While observations do seem to be
converging closer to unity than this, there are still many
uncertainties present.

For example, \nocite{Myers04}{Myers} {et~al.} (2004) found evidence for an extended
Sunyaev-Zel'dovich Effect in WMAP data out to about 1 degree from
centres of $z< 0.2$ clusters of galaxies, 
and they note that the contribution from clusters at $z > 0.2$ could 
significantly contaminate the $l=220$ first acoustic peak, so that 
the uncertainties stated in estimating $\Omtot \sim 1$, e.g. 
$\Delta \Omtot =0.02$ \nocite{WMAPSpergel}({Spergel} {et~al.} 2003), may be considerably 
underestimated due to sources of systematic error such as
the Sunyaev Zel'dovich Effect.

Hence, three-dimensional methods of detecting cosmic topology still
remain of interest, even though consistency with WMAP analyses would
be required for any serious 3-manifold candidates.

Given the many systematic effects, both due to observation and to 
physical aging of objects, in catalogues of extragalactic objects,
it is quite possible that a real signal could be present, but that 
the statistical properties of the catalogue would be insufficient
to establish statistical significance from general statistical 
properties of the catalogue.

Any additional tests which could either falsify or support a given
topological lensing hypothesis would, therefore, be useful.

In this paper, a simple property of topological lensing, 
which is only valid for the spherical case, is presented as such 
a test. 

{\em This is the necessary condition that the
mapping or isometry between the two sets of objects, of which 
one set is the topological image of the other, 
must be a whole number root of the identity in the covering space, 
${{}{S}}^3$.} This is because ${{}{S}}^3$ is finite, while ${\mathbb{R}}^3$
and ${{H}}^3$ are not.

\bfrevisedversion{
The 
covering space, ${{}{S}}^3$, ${\mathbb{R}}^3$ or ${{H}}^3$, 
(which can be physically thought of as the {\em apparent space})
relates to the 3-manifold, $M$, 
(which can be physically thought of as the {\em fundamental domain}
with glued faces), by $M = {{}{S}}^3/\Gamma$, 
${\mathbb{R}}^3/\Gamma$ or ${{H}}^3/\Gamma$ respectively, where 
$\Gamma$ is a group of isometries, which can be more generally called
a {\em holonomy group}.\footnote{See 
\protect\url{http://en.wikipedia.org/wiki/Holonomy} for 
a formal definition of holonomy group.}

The reason why the condition that the mapping must be a whole
number root of the identity is necessary in $S^3$ but not in flat or
negatively curved space can equivalently be understood as the fact
that the holonomy group $\Gamma$ is finite for $S^3$, but infinite in 
the other two cases.
}

Testing this property 
enables numerical falsification of a topological lensing hypothesis 
without needing to make any assumption on {\em which} spherical 3-manifold
would correspond to the isometry which has been found.
This is a practical advantage for observers interested in making 
simple analyses of observational data. 

Moreover, 
as mentioned by \nocite{GausSph01}{Gausmann} {et~al.} (2001) in eq.~(15) of their paper,
by embedding $S^3$ in ${\mathbb R}^4$, 
the isometry can be written as a combination
of two simultaneous rotations in orthogonal 2-planes in ${\mathbb R}^4$.
Hence, 
the rotations in both 2-planes also need to be {\em whole number roots
of the identity in ${\mathbb R}^2$}, i.e. their angles need to 
be whole number fractions of $2\pi$, since the covering space can only be
covered once.

This implies a simple and rapid test for testing whether or not
a numerically known isometry is a root of the identity.

To make this paper self-contained,
the complete algebraic formulae needed for these calculations, given 
the celestial positions and redshifts of corresponding objects, are
presented.

In \SSS\ref{s-crystal}, a reminder on the terminology 
regarding different types of pairs of topologically
lensed objects is presented. 

The condition that the isometry is a root of
the identity is presented algebraically and discussed
in \SSS\ref{s-root}.  
An abstract representation of the isometry is written 
in \SSS\ref{s-abstract-root-M}, while the 
four-dimensional matrix representations of pairs --- embedded in 
Euclidean 4-space, ${\mathbb{R}}^4$ --- 
and the resulting 
calculation of the isometry given the objects' sky positions and
redshifts, is presented in \SSS\ref{s-formulae}.
The condition that the isometry is a root of the identity is then
presented in a form easy to calculate in \SSS\ref{s-root-M}.

In \SSS\ref{s-root-angles}, the condition that the two rotation angles
of the isometry are whole number roots of $2\pi$ is presented.
In \SSS\ref{s-root-angles-ideal}, the ideal case in which the isometry
happens to be expressed in a convenient basis is presented.
In \SSS\ref{s-eigenplanes}, the more realistic case of an isometry expressed
in an arbitrary orthonormal basis is presented as an {\em eigenplane problem} 
(in analogy with eigenvectors):
an Earth or Solar System or Galaxy based coordinate system is unlikely to
be aligned with the eigenplanes of the isometry of comoving space --- in the 
case that the comoving space we live in is 
a multiply connected, spherical 3-manifold.

A method of finding the eigenplanes is presented in \SSS\ref{s-eigenplane-calc},
and a numerical implementation of this is shown in \SSS\ref{s-numerical}.

A summary is presented in \SSS\ref{s-conclu}.

For review papers on cosmic topology, see, 
e.g. \nocite{LaLu95,Lum98,Stark98,LR99}{Lachi\`eze-Rey} \& {Luminet} (1995); {Luminet} (1998); {Starkman} (1998); {Luminet} \& {Roukema} (1999); workshop proceedings 
are in \nocite{Stark98}{Starkman} (1998) and following articles,
and \nocite{BR99}{Blanl{\oe}il} \& {Roukema} (2000). 
For comparison and classification of different observational strategies,
see e.g. \nocite{ULL99b,LR99,Rouk02topclass,RG04}{Uzan} {et~al.} (1999a); {Luminet} \& {Roukema} (1999); {Roukema} (2002); {Rebou\c{c}as} \& {Gomero} (2004). 

\fpairtypes

\section{Definitions and calculation of the isometry} 
\subsection{Cosmic crystallography: local isometries (type I pairs) 
vs generator pairs (type II)}
\label{s-crystal}

The isometry between two images of a single region of physical
space yields two types of pairs of objects which can reveal 
the isometry, and uncorrelated pairs: 
\begin{list}{(\Roman{enumi})}{\usecounter{enumi}} 
\item
{\em local} or {\em type I} pairs which should occur
multiple times independently of curvature,
\item
{\em generator} or {\em type II} pairs, which only occur multiple
times for a 3-manifold in which there are Clifford translations, i.e.
in the flat and spherical cases,
\item 
uncorrelated pairs, which can be called ``type III'' pairs.
\end{list}

These are schematically illustrated in Fig.~\ref{f-pairtypes}.

Correspondingly, principles of detecting these pairs have been 
developed:
\begin{list}{(\Roman{enumi})}{\usecounter{enumi}} 
\item
The type I pairs can be collected, more generally,
as local $n$-tuplets \nocite{Rouk96}({Roukema} 1996),
or the two-point auto-correlation function 
of the two-point auto-correlation function, known as 
``collecting correlated pairs'' (CCP) \nocite{ULL99a, ULL99b}{Uzan} {et~al.} (1999b, 1999a) can be used.
\nocite{Gomero99a}{Gomero} {et~al.} (2002) noted these pairs as small deformations of the 
pair histogram, used for detecting type II pairs. 
\item 
In a pair separation histogram (PSH), type II pairs should,
ideally, show up as sharp spikes --- this is 
``cosmic crystallography'' \nocite{LaLu95,LLL96,Gomero99a}({Lachi\`eze-Rey} \& {Luminet} 1995; {Lehoucq} {et~al.} 1996; {Gomero} {et~al.} 2002). 
\end{list}

As in most of observational cosmology, observed catalogues of objects
are never as simple as could na\"{\i}vely be hoped for making 
cosmological measurements. It could realistically be the case
that a catalogue of objects contains a real topological signal, e.g. 
as in the right-hand plot of fig.~15 of \nocite{GausSph01}{Gausmann} {et~al.} (2001),
where $(\Omm=0.35,\Omega_\Lambda=0.75)$, i.e. $\Omtot = 1.1$, 
detected either thanks to collecting together type I pairs or
type II pairs, but that evolutionary and selection effects make
the signal of only weak statistical significance.

While it is certainly possible to simply ignore such isometries which
cannot be shown to be significant, it would good to be able to have
some test which relates to their immediate geometrical properties and
not to their membership of a statistical class.

This is the case presented in this paper: whether the suspected isometry 
is due to type I or type II pairs, if it is an isometry for the spherical
case, then there {\em does} exist a simple test capable of falsifying
the \bfrevisedversion{topological} lensing hypothesis.

Let us use the notation shown  in Fig.~\ref{f-pairtypes}, where only one
suspected realisation of the isometry $g$ is shown, mapping four points
in one image of the fundamental domain to four points in another
image:  
\begin{equation}
g: \{ {\bf a}_i \}_{i=1,4} \rightarrow  \{ {\bf b}_i \}_{i=1,4}.
\end{equation}

\bfrevisedversion{Although only three points are necessary to uniquely
define an isometry $g$, four points are necessary if we wish to use 
the embedding in four-dimensional euclidean space as discussed below 
in \SSS\ref{s-formulae}. Given empirical uncertainties, it is probably 
useful to have the extra information provided by the fourth pair, which 
in the absence of observational uncertainties and numerical errors, would
be redundant.} 
The case of more than four points is a generalisation of this, including
\bfrevisedversion{more} redundant information.

\subsection{Root of the identity in ${{}{S}}^3$}
\label{s-root}

\subsubsection{Abstract representation}
\label{s-abstract-root-M}

We can then write the first condition in the case of ${{}{S}}^3$ which 
must be satisfied if $g$ really is an isometry of the covering space
which generates a 3-manifold,
and not just a coincidence:
\begin{equation}
g^n( {\bf x} ) \equiv g ( g ( \ldots  g ( {\bf x} ) \ldots ) ) = I
\label{e-root-g}
\end{equation}
where $I$ is the identity mapping, must be true for some whole number 
$n \in {\mathbb{Z}}$.

Moreover, the value of $n$ should not be so high that $g^n$ ``goes 
past'' one loop of $2\pi$ around the hyper-sphere and evaluates to the
identity after tiling the whole covering space twice or more. In other
words, the tiling of ${{}{S}}^3$ by copies of the fundamental domain 
should only cover ${{}{S}}^3$ once.

To write this more formally, using a formalism which also enables
the expression  Eq.~(\ref{e-root-g}), 
in a way that is conceptually simple and easy to calculate,
it is useful to embed ${{}{S}}^3$ in four-dimensional 
euclidean space, ${\mathbb{R}}^4$.

\subsubsection{4-D representations in ${\mathbb{R}}^4$ and calculation
of the isometry}
\label{s-formulae}

By embedding ${{}{S}}^3$ in four-dimensional 
euclidean space, ${\mathbb{R}}^4$, the reader's intuition of 
${{}{S}}^2$ embedded in three-dimensional 
euclidean space, ${\mathbb{R}}^3$ can be used. 

The distance between two points 
(object locations)
in comoving space can then be thought of as an arc-length 
in ${\mathbb{R}}^4$, along the 3-surface ${{}{S}}^3$ \nocite{Rouk01-4D}(e.g., {Roukema} 2001).

Let us write the $j$-th points of the $i=1,2$ members of the
$j$-th pair of corresponding objects, i.e. where 
\begin{equation}
g(\bmath{a}_j) = \bmath{b}_j
\label{e-g}
\end{equation}
as
\begin{equation}
\bmath{a}_j
= 
\left(
\begin{array}{c}
x_{1j} \\
y_{1j} \\
z_{1j} \\
w_{1j} \\
\end{array} 
\right),  \;\;
\bmath{b}_j
= 
\left(
\begin{array}{c}
x_{2j} \\
y_{2j} \\
z_{2j} \\
w_{2j} \\
\end{array} 
\right), 
\end{equation}
evaluating these from, e.g. eq.~(33) of \nocite{Rouk01-4D}{Roukema} (2001), 
\bfrevisedversion{ where
the objects are located at celestial positions
$\alpha_{ij}, \delta_{ij}$ and redshifts $s_{ij}$, }
\begin{eqnarray}
\Omega_\kappa &\equiv& \Omm + \Omega_\Lambda -1 
\nonumber \\
 R &\equiv &
\begin{array}{lccl}
 {(c/ H_0)} { ( \Omega_\kappa)^{-0.5}  } 
\end{array}
\nonumber \\
\chi_{ij} 
&=& {c \over H_0} \int_{1/(1+s_{ij})}^1 
{ \mbox{\rm d}a \over a \sqrt{\Omm /a - \Omega_\kappa + 
\Omega_\Lambda a^2} }
\nonumber \\
\nonumber \\
x_{ij}&=&  R\; \sin (\chi_{ij}/R) \;\cos\delta_{ij} 
                           \;\cos\alpha_{ij} \nonumber \\
y_{ij} &=&  R\; \sin (\chi_{ij}/R) \;\cos\delta_{ij} 
                           \;\sin\alpha_{ij}  \nonumber \\ 
z_{ij} &=&  R\; \sin (\chi_{ij}/R) \;\sin\delta_{ij} \nonumber\\
w_{ij} &=& 
         \begin{array}{lccl}
R\; \cos (\chi_{ij}/R) \\
         \end{array}.
\end{eqnarray}
\bfrevisedversion{Since 
we are dealing with spherical spaces here, $\Omega_\kappa$ is positive.}

In ${\mathbb{R}}^4$, 
the isometry $g$ between $\bmath{a}_j$ and $\bmath{b}_j$ is 
a rotation about the origin $(0,0,0,0)$ --- ${{}{S}}^3$ is the 3-sphere 
(hypersphere) of 
radius $R$ centred on the origin, without loss of generality.
This rotation can be written as a $4\times4$ matrix $M$ of unity
determinant.

Since $M$ must map $\bmath{a}_j$ to $\bmath{b}_j$ for each $j$ 
(Eq.~\ref{e-g}), we have
\begin{equation}
M \; A = B
\label{e-MAB}
\end{equation}
where $A$ and $B$ are the matrices
\begin{equation}
A \equiv 
\left( \bmath{a}_1 \;  \bmath{a}_2 \;  \bmath{a}_3 \; 
    \bmath{a}_4 \right), \; 
B \equiv 
\left( \bmath{b}_1 \;  \bmath{b}_2 \;  \bmath{b}_3 \; 
    \bmath{b}_4 \right).
\end{equation} 

\bfrevisedversion{If the four-vectors 
in ${\mathbb{R}}^4$ of all four points $\bmath{a}_j$ (equivalently, 
of $\bmath{b}_j$) are linearly independent, 
then} $A$ is invertible, and multiplication by $A^{-1}$ 
from the right-hand side of Eq.~\ref{e-MAB} gives
\begin{equation}
M  = B\; A^{-1},  
\label{e-mba}
\end{equation}
which should be a matrix of nearly unity determinant, apart from
the positional uncertainties which are considered ``acceptable'' for
the calculation. Discussion of what precision is ``acceptable'' is
postponed to \SSS\ref{s-acceptable-errors} below.

Hence, $M$ can be calculated from the celestial positions and redshifts of the
two pairs of four corresponding objects, together with the values of the
curvature parameters for which the isometry was found.

$M$ is a matrix representation of the generator $g$.

\bfrevisedversion{However, even if all four points are distinct, their four-vectors
in ${\mathbb{R}}^4$ are not necessarily linearly independent (e.g. 3
vectors could be aligned in the same 2-plane). Moreover, if the
$\bmath{a}_j$ are nearly, but not quite, aligned, then the system 
could be ill-conditioned, i.e. be highly sensitive to small errors.
Any numerical application of this method should either ignore
choices of 4-tuplets which are not linearly independent enough or at
least flag them and warn the user.}

\subsubsection{Euclidean 4-D representation of the root 
of the identity constraint}
\label{s-root-M}

The constraint on the generator $g \equiv M$ 
in Eq.~(\ref{e-root-g}) can now be
rewritten
\begin{equation}
M^n = I.
\end{equation}

Clearly, na\"{\i}vely testing this numerically to arbitrarily large
$n$ would not be a practical way to test this. 
\bfrevisedversion{This is because in the
case of incorrect hypotheses, a computer making
numerically exact (rather than approximate) calculations would require
a (countably) infinite number of calculations in order to check that
no $n$ is large enough to yield $M^n=I$. Real computers make approximations
and are subject to numerical errors --- as $n$ gets bigger, these errors
would increase without limit and make the calculation meaningless at
some large value of $n$, unless the algorithm recalculated each successive
estimate of $M^n$ to higher and higher precision, at the expense of increasing
the computing time per calculation as $n$ gets higher, ensuring no possibility
of a convergent series for the total computing time requiring.}

\subsection{Two rotation angles as roots of the 2-D identity}
\label{s-root-angles}

\subsubsection{Constraint represented in a well-chosen basis}
\label{s-root-angles-ideal}

However, as described using a four-dimensional matrix representation
in eq.~(15) of \nocite{GausSph01}{Gausmann} {et~al.} (2001), there exists an 
orthonormal basis in which $M$ can be expressed in the 
form 
\begin{equation}
M = \left[ 
\begin{array}{cccc}
\cos\theta & -\sin\theta & 0 & 0 \\
\sin\theta & \cos\theta & 0 & 0 \\
0 & 0 & \cos\phi & -\sin\phi \\
0 & 0 &  \sin\phi & \cos\phi 
\end{array}
\right]
\label{e-m-twoplanes}
\end{equation}
where the rotation angles $\theta$ and $\phi$ clearly have to satisfy
\begin{equation}
   {2\pi \over \theta},
   {2\pi \over \phi} 
 \in {\mathbb{Z}}.
\label{e-anglesinZ}
\end{equation}
Writing the \bfrevisedversion{least common multiple} as 
\begin{equation}
 n \equiv \mbox{\rm LCM} \{ 
   {2\pi \over \theta},
   {2\pi \over \phi} \},
\end{equation}
we then have the smallest value $n$ such that $M^n = I$. If either
$   {2\pi \over \theta}$ or
$   {2\pi \over \phi} $
are not integers, then clearly $M$ is not a root of the identity.

\subsubsection{Eigenplanes}
\label{s-eigenplanes}

However, the orthonormal basis in ${\mathbb R}^4$ 
corresponding to a given astronomical coordinate system is unlikely,
in general, to be the basis in which $M$ already has this form.

The representation of $M$ in Eq.~(\ref{e-m-twoplanes}) has 
four orthonormal basis vectors.
Let us write these as
\begin{equation}
{\bf s},{\bf t}, {\bf u},{\bf v} \nonumber
\end{equation} 
so that 
\begin{eqnarray}
M {\bf s} &=& \cos\theta \;{\bf s} + \sin\theta \;{\bf t} \nonumber \\
M {\bf t} &=& -\sin\theta\;{\bf s}+ \cos\theta \;{\bf t} \nonumber \\
M {\bf u} &=& \cos\phi \;{\bf u} + \sin\phi \;{\bf v}  \nonumber \\
M {\bf v} &=& -\sin\phi \;{\bf u} + \cos\phi \;{\bf v}  
\end{eqnarray}

In analogy with eigenvectors and eigenvalues, we can call this 
an eigenplane problem, where:
\begin{equation}
\begin{array}{c}
M 
\left[ \begin{array}{c} {\bf s} \; {\bf t } \end{array} \right] 
= 
\left[ \begin{array}{c} {\bf s} \; {\bf t } \end{array} \right]
\Lambda_\theta , 
\end{array}
\;
\begin{array}{c}
M 
\left[ \begin{array}{c} {\bf u} \; {\bf v } \end{array} \right] 
= 
\left[ \begin{array}{c} {\bf u} \; {\bf v } \end{array} \right]
\Lambda_\phi, 
\end{array}
\label{e-eigenplanerot}
\end{equation}
and
\begin{equation}
\begin{array}{c}
\Lambda_\theta = \left[ 
\begin{array}{cc}
\cos\theta & -\sin\theta  \\
\sin\theta & \cos\theta 
\end{array}
\right], 
      \end{array} 
\;\;
\begin{array}{c}
\Lambda_\phi = \left[ 
\begin{array}{cc}
 \cos\phi & -\sin\phi \\
 \sin\phi & \cos\phi 
\end{array}
\right],
        \end{array}
\label{e-lambdaphi}
\end{equation}
i.e. the 1-dimensional 
\bfrevisedversion{(scalar) eigenvalue $\lambda$ of the traditional problem is 
replaced by a 2-dimensional eigenrotation $\Lambda_\theta$ or $\Lambda_\phi$.}

Just as there is freedom up to multiplication by a scalar for
the eigenvectors of an eigenvector problem, there is freedom up 
to rotation by an arbitrary (non-zero) rotation to find the 
eigenplane in the eigenplane problem. If 
$\left[ \begin{array}{c} {\bf s} \; {\bf t } \end{array} \right]$ is 
an orthonormal basis representing one eigenplane of $M$, with eigenrotation 
$\Lambda_\theta$, then
$\left[ \begin{array}{c} {\bf s} \; {\bf t } \end{array} \right] A $, 
where $A$ is an arbitrary 2-dimensional rotation matrix,
is also an orthonormal basis for the same eigenplane and same eigenrotation:
\begin{equation}
\begin{array}{c}
M 
\left[ \begin{array}{c} {\bf s} \; {\bf t } \end{array} \right] 
A 
= 
\left[ \begin{array}{c} {\bf s} \; {\bf t } \end{array} \right]
A 
\Lambda_\theta. 
\end{array}
\label{e-freerotation}
\end{equation}

In the particular case of interest here, there should exist 
two orthogonal eigenplanes
in order for the isometry to correspond
to a holonomy transformation, i.e. for it to be of interest for generating a
3-manifold as a quotient space of
the 3-sphere centred at the origin of ${\mathbb R}^4$.

\subsubsection{Calculating the eigenplanes}
\label{s-eigenplane-calc}

How do we find the basis vectors, now grouped into two pairs, 
$({\bf s}, {\bf t}), ({\bf u}, {\bf v})$ 
spanning these two 2-planes respectively?

Since eigenvalue problems are normally solved using an iterative 
algorithm, it seems natural to develop a practical algorithm for solving
the eigenplane problem.

\newcommand\minbasisprobablynotneeded{
Let us write $M$, in the {\em known} basis, as
\begin{equation}
M = \left[ 
\begin{array}{cccc}
a & b& c & d \\
e & f& g & h \\
i & j& k & l \\
m & n& o & p 
\end{array}
\right]
\end{equation}
where $a$--$p$ are calculated from Eq.~(\ref{e-mba}).


Consider two arbitrary, orthonormal 
unit vectors in this known basis, e.g. two of the basis vectors 
\begin{equation}
\bmath{e}_1
= 
\left(
\begin{array}{c} 1 \\ 0 \\ 0 \\ 0
\end{array} 
\right) \;\;
\bmath{e}_2
= 
\left(
\begin{array}{c} 0 \\ 1 \\ 0 \\ 0
\end{array} 
\right).
\end{equation}
}


{\em Suppose} that we already have one vector in 
one of the two eigenplanes $P_1$ or $P_2$. Then, without loss
of generality (wlog), we can write this
\begin{equation}
{\bf s} \in P_1 \nonumber
\end{equation}
and ${\bf s}$ and $M {\bf s}$ can be used to construct a second
vector in the same 2-plane, yielding a pair of orthonormal vectors
$({\bf s}, {\bf t})$:
\begin{eqnarray}
{\bf t} &\equiv& 
  { M {\bf s} - ({\bf s} \cdot M {\bf s}) \; {\bf s}  \over
 \sqrt{ 1 - ({\bf s} \cdot M {\bf s})^2 } },
\label{e-planeone-trivial}
\end{eqnarray}
where $\cdot$ is the inner product on ${\mathbb R}^4$.

\bfrevisedversion{ 
It is clear that ${\bf s}$ and ${\bf t}$ are orthogonal, since
\begin{eqnarray}
{\bf s} \cdot {\bf t} &=&
{  {\bf s} \cdot  M {\bf s} - 
{\bf s} \cdot ({\bf s} \cdot M {\bf s}) \; {\bf s}  \over
 \sqrt{ 1 - ({\bf s} \cdot M {\bf s})^2 } } \nonumber \\
&=&
{ ( {\bf s} \cdot  M {\bf s})  
( 1- {\bf s} \cdot {\bf s}) \over
 \sqrt{ 1 - ({\bf s} \cdot M {\bf s})^2 } } 
\label{e-proof-s-t-ortho}
\end{eqnarray}
and $1 - {\bf s} \cdot {\bf s} = 0 $ since ${\bf s}$ is a unit vector.
}


Next, we need a vector which is not in $P_1$. 
Consider three (orthonormal) basis vectors, 
${\bf e}_1,  {\bf e}_2,  {\bf e}_3$,
in the known basis, \bfrevisedversion{i.e. the basis in which 
$\bmath{a}_j$ and $\bmath{b}_j$ are calculated.}
Since these are orthogonal to
one another, at most two of them can be in $P_1$.  
Choose one of these outside of $P_1,$ let us call it ${\bf e}_1$ wlog.
Then, 
\begin{eqnarray}
{\bf u} &\equiv& {\bf e}_1 - 
({\bf e}_1 \cdot {\bf s}) \; {\bf s} - 
({\bf e}_1 \cdot {\bf t}) \; {\bf t}
 \nonumber \\
{\bf v} &\equiv& 
  { M {\bf u} - ({\bf u} \cdot M {\bf u}) \; {\bf u}  \over
 \sqrt{ 1 - ({\bf u} \cdot M {\bf u})^2 } } ,
\label{e-planetwo-trivial}
\end{eqnarray}
are an orthonormal pair spanning the second plane, $P_2$.

We then have $\theta$ and $\phi$ from
\begin{eqnarray}
\theta &=& \cos^{-1} ({\bf s} \cdot M {\bf s}) \nonumber \\
\phi &=& \cos^{-1} ({\bf u} \cdot M {\bf u})  
\label{e-solution-specialcase}
\end{eqnarray}

Hence, finding a solution for 
$({\bf s}, {\bf t}), ({\bf u}, {\bf v})$ 
reduces to finding at least one vector ${\bf s}$ in one 
of the two planes $P_1, P_2$.

The intersection of $P_1$ (or $P_2$) with $S^3$ is a great circle, centred on the
origin. This must intersect somewhere with the 2-sphere defined by
\begin{eqnarray}
S^2_{123} &\equiv&
\mbox{span}\{ {\bf e}_1,  {\bf e}_2,  {\bf e}_3 \} 
\cap
S^3.
\end{eqnarray}

Hence, it is sufficient to search through $S^2_{123}$ looking 
for at least one vector which lies in $P_1$ or $P_2$. In fact, 
it is sufficient to search through half of this region, since polarity
is unimportant.

To write down 
the condition for a vector ${\bf s}$ to be in one of the two eigenplanes, 
note that the three vectors
\begin{equation}
{\bf s}, M {\bf s}, M^2  {\bf s},
\end{equation} 
must all lie in the same eigenplane. This eigenplane intersects with 
$S^3$ in a great circle, i.e. rotating twice corresponds to a single rotation
by twice the angle of the original rotation.
We can write this condition as 
\begin{equation}
\cos^{-1} ({\bf s} \cdot M^2 {\bf s}) =
2 \cos^{-1} ({\bf s} \cdot M {\bf s}) 
\label{e-condition-one}
\end{equation}
for a rotation angle 
\begin{equation}
\theta \equiv \cos^{-1} ({\bf s} \cdot M {\bf s}) 
\le \pi/2
\label{e-theta-one}
\end{equation}
or 
\begin{equation}
2\pi - \cos^{-1} ({\bf s} \cdot M^2 {\bf s}) =
2 \cos^{-1} ({\bf s} \cdot M {\bf s}),
\label{e-condition-two}
\end{equation}
for 
\begin{equation}
\pi/2 \le \theta \equiv \cos^{-1} ({\bf s} \cdot M {\bf s}) 
\le \pi.
\label{e-theta-two}
\end{equation}


Hence, an iterative search to sucessively preciser resolution
through $S^2_{123}$ to find a vector ${\bf s}$ satisfying
either Eq.~(\ref{e-condition-one}) or Eq.~(\ref{e-condition-two}) 
to the desired numerical precision yields
one basis vector ${\bf s}$ of the eigenplanes, and 
Eqs.~(\ref{e-planeone-trivial}), (\ref{e-planetwo-trivial}) yield the
other three basis vectors. 

\bfrevisedversion{Tests using the {\tt eigenplane} package 
(\SSS\ref{s-numerical}) suggest that the number of steps $n$ to 
reach a precision in radians of $\Delta \theta$ is 
\begin{equation}
n = 2 - 4 \log_{10} \Delta \theta,
\end{equation}
i.e. for  $\Delta \theta \sim 10^{-4}$--$10^{-3},$ about 14--18 iterations 
are sufficient.}

The rotation angles corresponding to the isometry 
are those in Eq.~(\ref{e-solution-specialcase}).

If these angles do not satisfy Eq.~\ref{e-anglesinZ}, then the isometry is
not a root of the identity and the topological lensing hypothesis is false.

\bfrevisedversion{
\subsection{Numerical aspects}
}

\subsubsection{Numerical implementation}
\label{s-numerical}

The package {\tt eigenplane}, 
which is a free software (GNU GPL licence) implentation of an iteration 
algorithm to generate the two rotation angles $\theta, \phi$ 
(and as a side product,
a choice of four orthonormal vectors 
$({\bf s}, {\bf t}), ({\bf u}, {\bf v})$ 
defining the two eigenplanes),
given an input isometry $M$, has been prepared and is
available for download at 
\url{http://cosmo.torun.pl/GPLdownload/eigen/}.\footnote{At the time
of proofchecking this article, the current version of the package 
is {\tt eigenplane-0.2.3}. Users are welcome to add features, correct bugs 
and distribute modified versions.}

The package is self-contained, apart from requiring fortran and C
compilers and the public domain linear algebra library {\tt blas}.
Installation (help in {\tt README} and {\tt INSTALL} files) 
is by the standard {\tt ./configure; make; make install}
sequence, including standard options such as a non-root user
installation directory via {\tt --prefix=PREFIX}.
Following installation, help is available with the
commands {\tt eigenplane --help} and {\tt info eigenplane}.


\bfrevisedversion{

\subsubsection{What are ``acceptable'' levels of errors in the positions?}
\label{s-acceptable-errors}

As was discussed, e.g. in \nocite{Rouk96}{Roukema} (1996) and later papers, 
these consist, in principle, both of ``measurement'' errors --- to what extent
the observed values of angular positions 
$\alpha_{ij}, \delta_{ij}$ and redshift $s_{ij}$\footnote{The usual variable
$z$ is used here for positions in $\mathbb{R}^4$, so we write $s$ (red-``shift'') 
instead.} incorrectly represent the true position if peculiar velocity 
is ignored --- and of ``movement'' error --- to what extent the objects
move relative to the comoving reference frame between the different epochs
(redshifts) at which they are observed.

In practice, the precision on the angular positions is negligible 
compared to all other errors. The former are almost always more accurate than
an arcsecond, which at a redshift of $s=2$ and 
$(\Omm=0.35, \Omega_\Lambda=0.75)$ corresponds to a tangential error of
better than 0.02{\hMpc}. 

The measurement error in the redshifts, e.g. of quasars, {\em can} usually
be obtained to $\Delta s < 0.001$, but in big catalogues is more often
$\Delta s < 0.01$.  Again at a redshift $s=2$ and 
$(\Omm=0.35, \Omega_\Lambda=0.75)$, these two errors lead to distance
errors of about 1{\hMpc} and 10{\hMpc} respectively, at least 50 times larger
than that induced by any angular error. 

The movement error between observations at different redshifts depends
on how different the redshifts are, and on hypotheses of structure formation
within the general model of gravitational collapse from linear perturbations.
The objects most likely to be observed are the brighter ones, which tend
to lie in the heavier potential wells, which can be expected to move more
slowly relative to the comoving reference frame. 

If we estimate a maximum for the peculiar velocity as a mean of 400~km/s then
between two substantially different redshifts, e.g. differing by 4~Gyr, the
total displacement (relative to the comoving reference frame) is about 
1.6{\hMpc}. Moreover, simulations suggest that some objects may ``stream''
from voids towards filaments and along filaments towards the deepest potential
wells in the cosmic web of gravitationally bound structure --- i.e. movement
by up to a few megaparsecs over a big fraction of a Hubble time is realistic for
objects forming some distance away from the biggest clusters of galaxies.

So likely errors are in the range 1--10{\hMpc}, both from redshift measurements
and from possible movement relative to the comoving frame.

How much do these affect testing an isometry as a root of the identity?

Again, for $(\Omm=0.35, \Omega_\Lambda=0.75)$, the curvature radius is
$R \approx 10${\hGpc}, so 
the errors in three-dimensional position, i.e. within
the 3-surface $S^3$, are at the
level of about $10^{-4}$ to $10^{-3}$ curvature radii, i.e. about
$10^{-4}$ to $10^{-3}$ radians.
If these lead to a similar error in the estimates
of the angles $\theta$ and $\phi$, and the magnitude of 
$\mbox{\rm max} (\theta,\phi)$ is a big
fraction of the distance to the surface of last scattering, 
e.g. $\ge 5${\hGpc} $\sim 0.5$ radians, then the fractional (relative) errors in 
$\mbox{\rm max}({2\pi \over \theta},   {2\pi \over \phi})$ 
should be of the same order of magnitude,
about $10^{-4}$ to  $10^{-3}$. 
A reasonable lower bound could be placed on
$\mbox{\rm max} (\theta,\phi)$, for a given
$(\Omega_{\mbox{\rm m}},\Omega_\Lambda)$ pair,
in order for the isometry to be realistic
(e.g. given cosmic topology constraints from the cosmic microwave background),
removing isometries where
$\mbox{\rm max}(\theta, \phi)$ is too small.
So, for holonomy groups of order up to $\sim 10^1$, the probability of 
false (chance) isometries yielding integer solutions is about 
$10^{-3}$ to  $10^{-2}$. For larger orders, the probability of false
solutions necessarily becomes higher. 

It remains possible that the smaller angle, 
$\mbox{\rm min}(\phi, \theta)$, could be quite small and would have a 
larger relative error (e.g. $10^{-2}$--$10^{-1}$), so in many cases would
constitute a weaker test of candidate isometries.

If the values 
$(\Omm,\Omega_\Lambda)$ are decreased to approach $\Omm+\Omega_\Lambda=1$,
then the curvature radius $R$ increases rapidly but the distance to the surface
of last scattering only changes slightly, so $\mbox{\rm max} (\theta,\phi)$
decreases. This causes $\mbox{\rm max}({2\pi \over \theta},   {2\pi \over \phi})$
to increase rapidly, so even while the relative error will not increase,
the {\em absolute} error will increase to above $\pm 1$, in which case
integer solutions will necessarily be found for chance isometries. 

However, as pointed out by \nocite{Mota05}{Mota} {et~al.} (2005), as the curvature radius 
becomes larger and larger, observationally detectable isometries
in spherical (and hyperbolic) 
3-manifolds become closer and closer to those for the flat case, in which 
case methods used in the flat case are likely to become preferable.
}

\section{Discussion and conclusion}
\label{s-conclu}

Of course, checking that an isometry is a root of the identity
is only a {\em necessary} condition for the topological lensing 
hypothesis to be correct, it is not {\em sufficient}. Moreover,
even if it genuinely is a case of topological lensing, the isometry
might, in principle, not be an isometry between {\em adjacent} copies
of the fundamental domain --- or in other words, the group it generates
as the $n$-th root of the identity might only be a sub-group of the
full holonomy group \nocite{Weeks03}({Weeks} {et~al.} 2003).

Note that there is no point checking the Poincar\'e dodecahedral 
hypothesis of \nocite{LumNat03}{Luminet} {et~al.} (2003) and \nocite{RLCMB04}{Roukema} {et~al.} (2004) by this method, since
the isometries there are already known to be 
roots of the identity; the assumption of a particular 
class of 3-manifolds was an input assumption.

The situations where this test can be useful are those where a candidate
isometry is obtained for the spherical case {\em without} having assumed any
particular 3-manifold, but only having assumed that curvature is positive.

\bfrevisedversion{How can these candidate isometries be obtained?  In
other words, what are the known systematic methods for exploring a
catalogue of objects extending to high redshifts in order to find a
suitable pair of 4-tuplets of corresponding objects with a corresponding $g$
which should be tested?

In principle, the method of finding local $n$-tuplets of
\nocite{Rouk96}{Roukema} (1996) provides the answer: compare all ``local'' (less than a
few hundred Mpc) ``configurations'' of objects against all other
``local clusters'', each time checking whether the mapping $f$ between the two
4-tuplets is an isometry $g$ or not.  In practice, optimising the algorithm for
application to large, recent catalogues is a non-trivial task, since
the number of $n$-tuplets climbs rapidly as the number of objects
increases: the na\"{\i}ve approach with a large catalogue quickly
becomes impractical even on the latest supercomputers, though some suggestions
for shortcuts to the algorithm are made in \nocite{Rouk96}{Roukema} (1996).

A related approach might be to use the 
``collecting correlated pairs'' method (CCP) \nocite{ULL99a, ULL99b,Gomero99a}({Uzan} {et~al.} 1999b, 1999a; {Gomero} {et~al.} 2002), 
to first obtain a list of 2-tuplets (pairs) ``most likely to be matched 
type I pairs'', 
then try to combine these into 4-tuplets, test whether or not the 
mappings $f$ are isometries $g$, and finally test whether any ``best''
isometry $g$ is a root of the identity, using the algorithm presented here. 
The definition of the ``most likely to be matched type I pairs'' might
first be some way of choosing the highest signal-to-noise ratios as a function of 
the curvature parameters $(\Omm, \Omega_\Lambda)$ (as is recommended 
in \nocite{ULL99a,ULL99b}{Uzan} {et~al.} (1999b, 1999a)), and then for each choice of 
$(\Omm, \Omega_\Lambda)$, choose the bins in the pair separation histogram
with the highest numbers of pairs as the ``most likely to be matched
type I pairs''.

Whereas \nocite{ULL99a,ULL99b}{Uzan} {et~al.} (1999b, 1999a) hope for a strong signal, use of the
algorithm presented here could potentially enable detection of a 
candidate 3-manifold even if the signal is weak.

Type II pairs --- generator pairs --- detected as high spikes 
in a pair separation histogram (PSH) \nocite{LaLu95,LLL96,Gomero99a}({Lachi\`eze-Rey} \& {Luminet} 1995; {Lehoucq} {et~al.} 1996; {Gomero} {et~al.} 2002), 
could also be used to generate pairs of 4-tuplets and isometries $g$.  
}

\bfrevisedversion{Another caveat is that 
this test necessarily depends on the choice
of curvature parameters $(\Omm, \Omega_\Lambda)$. 

If an isometry $g$ is known numerically 
based on, e.g. a pair of 4-tuplets of observed objects for a 
given choice of $(\Omm, \Omega_\Lambda)$, then it is fairly likely that 
the same pair of 4-tuplets will be approximately isometric for 
nearby values of $(\Omm, \Omega_\Lambda)$, especially if the 4-tuplets
are local 4-tuplets (composed of type I pairs rather than type II pairs).
So, the range of $(\Omm, \Omega_\Lambda)$ for which the mapping
between a given pair of 4-tuplets remains an isometry
needs to be tested, unless external constraints 
on $(\Omm, \Omega_\Lambda)$ are considered acceptable. 
This (unfortunately) increases the chance that a false
(chance) isometry will yield a valid root of the identity.}

\section*{Acknowledgments}

The author thanks an 
anonymous referee for many very constructive
comments.



\subm{\clearpage}

\nice{
%

}

\end{document}